\documentclass[%
 aip,
 jmp,%
 amsmath,amssymb,
 superscriptaddress,
 reprint,%
 nofootinbib
]{revtex4}
\usepackage{amsmath}
\usepackage{stmaryrd}
\usepackage{color}
\usepackage{array}
\usepackage{enumitem}
\usepackage{mathpazo}
\usepackage{setspace}
\usepackage{bm}
\usepackage{amssymb}
\usepackage{amsthm}
\usepackage{mathtools}
\usepackage{multirow}
\usepackage{cases}
\usepackage[colorlinks=true,linkcolor=blue,citecolor=blue,pdfauthor={ },pdftitle={ },pdfsubject={ },pdfkeywords={ }]{hyperref}
\usepackage{pgfplots}
\usepackage{lipsum}
\usepackage{youngtab}
\usepackage{booktabs}
\begin{document}
\title{Understanding the Optical Theorem of Scattering: Scattering Surface Area against Scattering Cross Section, an Example with Ellipsoidal Scattering}

\author{Youning Li}%
\affiliation{College of Science, China Agricultural University, Beijing, 100080, People's Republic of China}
%

%
%
%
%
%

\date{\today}

\begin{abstract}
In this paper, we propose using the scattering surface area rather than the scattering cross section to characterize the scattering behavior of ellipsoidal rigid bodies. We examined the scattering behavior of ellipsoidal rigid bodies, focusing on the relationship between their surface area and total scattering cross-section. Building on the foundational work of Carson Flammer, we utilize the spheroidal coordinate system to derive solutions for both prolate and oblate spheroids. Our analysis reveals that under the long-wavelength approximation, the total scattering cross-section is equivalent to the surface area of the ellipsoid, a relationship that holds true for both small and moderate eccentricities. This finding extends the established optical theorem, previously validated for spherical bodies, to more complex geometries.
\end{abstract}

\maketitle

\section{Introduction}

Scattering phenomena play a crucial role in various fields of physics, including optics, acoustics, and electromagnetic theory\cite{newton2013scattering,faran1951sound,balanis2012advanced}.
Understanding how different shapes and materials interact with incident waves is essential for applications ranging from radar technology to nuclear reaction\cite{hulst1981light,waterman1971symmetry,knott2004radar,fesbach1992theoretical,jin2015theory}.
Among the diverse geometries studied, ellipsoidal shapes are of particular interest due to their prevalence in natural and engineered systems, such as particles in colloidal suspensions and biological cells\cite{asano1975light,gouesbet1982scattering,wriedt2009light,wyatt1968differential}.

Historically, the study of scattering by rigid bodies can be traced back to the work of Mie in the early 20th century\cite{mie1908beitrage}, who provided a comprehensive solution for the scattering of light by spherical particles. Mie's theory laid the groundwork for understanding how the size, shape, and composition of particles influence scattering behavior\cite{kerker1969scattering}. Following this, significant progress was made in deriving solutions for more complex geometries, including spheroidal shapes\cite{logan1965survey,stevenson1953solution,asano1980light}, which are characterized by their axial symmetry and are particularly relevant for assessing scattering of waves in non-uniform media\cite{hulst1981light}. In recent years, the study of complex scattering geometries has attracted growing attention in various fields. For instance, in nuclear physics, Xiao et al.\cite{zhu2020coexistence,ou2020orientation} have investigated the elastic scattering of deformed nuclei using coupled-channels theory and microscopic analysis. Although their research focuses on a different physical system, the theoretical methods and computational techniques employed in handling complex scattering geometries, such as deformed nuclei, provide valuable insights and inspiration for the study of ellipsoidal scatterers.

The spheroidal coordinate system offers a powerful framework for solving scattering problems involving ellipsoidal geometries. This coordinate system facilitates the mathematical treatment of prolate and oblate spheroids, allowing for a more nuanced understanding of how these shapes scatter incident waves. A notable contribution to this field is the work of Carson Flammer\cite{flammer2014spheroidal,flammer1953vector}, who extensively explored the scattering properties of spheroidal geometries. Flammer's results have provided a solid foundation for the mathematical formulation of scattering problems, particularly in establishing the relationships between various spheroidal functions and their scattering characteristics.

A key result in scattering theory is the relationship established for rigid spherical bodies, wherein the total scattering cross-section in the long-wavelength limit is equivalent to the surface area of the sphere, and four times of the cross section. At short-wave limit, the scattering cross section is equal to $2\pi a^2$, which is half of the surface area of a sphere. This can be understood that at short-wavelength limit, the wave can only reach the half-sphere facing the wave. Therefore, it is equal to $2\pi a^2$. In long-wavelength, the wave can reach all the surface areas, thereby $4\pi a^2$. This principle, often referred to as the “optical theorem” or “optical scattering equivalence,” provides a foundational understanding of wave interactions with spherical shapes. However, the extension of this principle to ellipsoidal shapes, particularly for moderate eccentricities, has not been thoroughly investigated.

In this work, we focus on the scattering behavior of ellipsoids, specifically examining the relationship between their surface area and total scattering cross-section. A general derivation of the equivalence of scattering cross section with the surface area is not possible at present. However, we can use ellipsoids at the long-wavelength limit to test our assumption. Utilizing the long-wavelength approximation, we aim to determine whether the equivalence between surface area and scattering cross-section is valid for ellipsoids with varying eccentricities. Our findings indicate that this relationship holds true, providing valuable insights into the fundamental nature of scattering processes across different geometrical configurations. This paper presents our theoretical framework, calculations, and implications of these results, ultimately contributing to a deeper understanding of scattering phenomena in complex geometries and suggesting avenues for future research in this area.

\section{Spheroidal Differential Equations}
The prolate and oblate spheroid coordinate systems are derived by rotating the two-dimensional elliptic coordinate system around the major and minor axes of the ellipses, respectively. Typically, the $z$-axis is chosen as the axis of revolution, and the interfocal distance is denoted by $d$.

The prolate spheroidal coordinates are related to the three-dimensional Cartesian coordinates as follows:
\begin{subequations}\label{prolate coordinates}
\begin{align}
x &= \frac{d}{2} \sqrt{(1-\eta^2)(\xi^2-1)} \cos \phi, \\
y &= \frac{d}{2} \sqrt{(1-\eta^2)(\xi^2-1)} \sin \phi, \\
z &= \frac{d}{2} \eta \xi,
\end{align}
\end{subequations}
where $\xi \geq 1$, $-1 \leq \eta \leq 1$, and $0 \leq \phi < 2\pi$.

For the oblate spheroidal system, the relationships are similar:
\begin{subequations}\label{oblate coordinates}
\begin{align}
x &= \frac{d}{2} \sqrt{(1-\eta^2)(\xi^2+1)} \cos \phi, \\
y &= \frac{d}{2} \sqrt{(1-\eta^2)(\xi^2+1)} \sin \phi, \\
z &= \frac{d}{2} \eta \xi.
\end{align}
\end{subequations}
where $\xi \geq 0$, $-1 \leq \eta \leq 1$, and $0 \leq \phi < 2\pi$.

In the prolate spheroidal system, the surface defined by $\xi=\textrm{constant}>1$ represents an elongated ellipsoid of revolution, with a major axis of length $d\xi$ and a minor axis of length $d\sqrt{\xi^2-1}$. The degenerate surface $\xi=1$ corresponds to the straight line along the $z$-axis from $z=-0.5d$ to $z=0.5d$. The surface where $|\eta|=\textrm{constant} < 1$ is a hyperboloid of revolution with two sheets, characterized by an asymptotic cone whose generating line passes through the origin and is inclined at the angle $\theta=\arccos \eta$ to the $z$-axis. The degenerate surface where $|\eta|=1$ is part of the $z$-axis for which $|z| > 0.5d$. The surface defined by $\phi=\textrm{constant}$ is a plane through the $z$-axis, forming an angle $\phi$ with the $x,z$-plane.

In the oblate spheroidal system, the surface $\xi=\textrm{constant} > 0$ denotes a flattened ellipsoid of revolution, with a major axis of length $d\sqrt{\xi^2+1}$ and a minor axis of length $d\xi$. The surface $\xi=0$ corresponds to a circular disk of radius $a=0.5d$, lying in the $x,y$-plane and centered at the origin. The surface where $|\eta|=\textrm{constant} < 1$ is a hyperboloid of revolution with one sheet, characterized by an asymptotic cone whose generating line passes through the origin and is inclined at the angle $\theta=\arccos \eta$ to the $z$-axis. The degenerate surface where $|\eta|=1$ corresponds to the $z$-axis. The surface $\eta=0$ represents the $x,y$-plane, excluding the circular disk at $\xi=0$. The surface defined by $\phi=\textrm{constant}$ is a plane through the $z$-axis, forming an angle $\phi$ with the $x,z$-plane.

In the limit when the interfocal distance $d$ goes to $0$, both the prolate and the oblate spheroidal reduce to spherical coordinate system. For finite $d$, the surface $\xi=$ constant be becomes spherical as $\xi$ approaches infinity in each case. Therefore, when $\xi\rightarrow \infty$,
\begin{equation}\label{limit}
0.5d\xi\rightarrow r,\qquad \eta\rightarrow \cos \theta,
\end{equation}
where the $\theta$ and $r$ are the usual spherical coordinates.

The scattering problem of a plane wave by a rigid rotating ellipsoid reads:
\begin{equation}
    (\nabla^2 + k^2 - U(\bm{r}))\psi(\bm{r}) = 0,
\end{equation}
with the potential given by
\[
U(\bm{r}) =
\begin{cases}
\infty & \xi \leq \epsilon^{-1}, \\
0 & \xi > \epsilon^{-1},
\end{cases}
\]
where $\epsilon$ is the eccentricity.

In the prolate and oblate coordinate systems, the separation of variables respectively give:
\begin{subequations}
\begin{align}
\left[\frac{\partial}{\partial \eta}(1-\eta^2)\frac{\partial}{\partial \eta} + \frac{\partial}{\partial \xi}(\xi^2-1)\frac{\partial}{\partial \xi} + \frac{\xi^2-\eta^2}{(\xi^2-1)(1-\eta^2)} \frac{\partial^2}{\partial \phi^2} + c^2 (\xi^2 - \eta^2) \right] \psi &= 0,\label{prolate sep}\\
\left[\frac{\partial}{\partial \eta}(1-\eta^2)\frac{\partial}{\partial \eta} + \frac{\partial}{\partial \xi}(\xi^2+1)\frac{\partial}{\partial \xi} + \frac{\xi^2+\eta^2}{(\xi^2+1)(1-\eta^2)} \frac{\partial^2}{\partial \phi^2} + c^2 (\xi^2 + \eta^2) \right] \psi &= 0,\label{oblate sep}
\end{align}
\end{subequations}
where $c = \frac{1}{2} k d$.
It is clear that Eq(\ref{oblate sep}) could be obtained from Eq(\ref{prolate sep}) by the transformation
\begin{equation}\label{trans}
  \xi\rightarrow\pm i \xi,\qquad c\rightarrow \mp ic.
\end{equation}
In the following, we shall firstly calculate and analysis in the prolate coordinates and then transform to the oblate case.

The solution to Eq(\ref{prolate sep}) is:
\begin{equation}\label{stand wave solution}
\psi_{mn} = S_{mn}(c, \eta) R_{mn}^{(j)}(c\xi) \cos \phi / \sin \phi,
\end{equation}
where $S_{mn}(c, \eta)$ and $R_{mn}^{(j)}(c\xi)$ can be expanded using associated Legendre functions and Bessel functions, respectively:
\begin{equation}\label{angle}
S_{mn}(c, \eta) = {\sum\limits_{r=0,1}^{\infty}} \!\! ' d_r^{mn}(c) P_{m+r}^m (\eta),
\end{equation}
\begin{equation}\label{radius}
R_{mn}^{(j)}(c\xi) = \frac{1}{{\sum\limits_{r=0,1}^{\infty}} \!\! ' d_r^{mn}(c) \frac{(2m+r)!}{r!}} \left( \frac{\xi^2 - 1}{\xi^2} \right)^{1/2 m} \sum_{r=0,1}^{\infty} i^{(r+m-n)} d_{mn}^r (c) \frac{(2m+r)!}{r!} J_{m+r}^{(j)} (c\xi),
\end{equation}
where
\[
J_l^{(1)}(r) = j_l (r),\quad J_l^{(2)}(r) = n_l (r),\quad J_l^{(3)}(r) = h_l^{(1)} (r),\quad J_l^{(4)}(r) = h_l^{(2)} (r).
\]
The prime on the summation indicates that when $n-m$ is even, $r$ sums over all even numbers, and when $n-m$ is odd, $r$ sums over all odd numbers. The coefficient $d^{mn}_r$ must satisfied a series of recursion second order differential equations, due to the recursion formulas for the associated Legendre functions, and the solution of these $d^{mn}_r$ should guarantee that $d^{mn}_{r+2}/d^{mn}_{r}$ goes to zero fast enough to make sure the summation to be convergent.
The reason for adding that coefficient in front of $R_{mn}^{(j)}(c\xi)$ is to ensure that as $c\xi \to \infty$, $R_{mn}^{(j)}(c\xi)$ has the following asymptotic forms, which resemble the usual asymptotic forms of the Bessel functions:
\[
R_{mn}^{(1)}(c\xi) \xrightarrow{c\xi \to \infty} \frac{1}{c\xi} \sin (c\xi - n\pi/2)
\]
\[
R_{mn}^{(2)}(c\xi) \xrightarrow{c\xi \to \infty} -\frac{1}{c\xi} \cos (c\xi - n\pi/2)
\]
\[
R_{mn}^{(3,4)}(c\xi) \xrightarrow{c\xi \to \infty} \frac{1}{c\xi} \exp \left( \pm i(c\xi - n\pi/2) \right)
\]
Therefore, a general solution to the initial wave equation Eq(\ref{prolate sep}) should be
\begin{equation}\label{mathematical solution}
\psi_{\text{mathematics}} = \sum_{n,m} A_{mn} S_{mn}(c, \eta) \left[ \cos \gamma_{mn} R_{mn}^{(1)}(c\xi) - \sin \gamma_{mn} R_{mn}^{(2)}(c\xi) \right] \cos m \phi,
\end{equation}
where $\gamma_{mn}$ represents a constant related to $m$ and $n$ but independent of $\eta$ or $\xi$.
Since the wave function is zero inside the ellipsoid, we obtain an equation for $\gamma_{mn}$:
\begin{equation}\label{gamma mn}
\tan (\gamma_{mn}) = \frac{R_{mn}^{(1)}(c\epsilon^{-1})}{R_{mn}^{(2)}(c\epsilon^{-1})} = \frac{\sum_{r}' i^{(r+m-n)} d_r^{mn}(c) \frac{(2m+r)!}{r!} j_{m+r}(c\epsilon^{-1})}{\sum_{r}' i^{(r+m-n)} d_r^{mn}(c) \frac{(2m+r)!}{r!} n_{m+r}(c\epsilon^{-1})}.
\end{equation}
The asymptotic form of $\psi_{\text{mathematics}}$ as $\xi$ goes to infinity is:
\begin{equation}\label{math asym}
\psi_{\text{mathematics}} = \sum_{m,n} \frac{2i^n (2 - \delta_{0m})}{N_{mn}} A'_{mn} S_{mn}(c, \cos \theta) \frac{\sin (kr - n\pi/2 + \delta_{mn})}{kr} \cos m \phi.
\end{equation}

\section{Scattering Section}
In a scattering problem, we assume the incident wave is a plane wave, whose propagation direction is defined by the angles $(\theta_0, \phi_0)$, while the scattered wave is an azimuth-dependent spherical wave observed at infinity. Thus, from a physical perspective, the wave function $\psi(\bm{r})$ can be expressed as a combination:

\begin{equation}\label{physical solution}
\psi_{\text{physics}} = e^{i \bm{k} \cdot \bm{r}} + \frac{f(\theta, \phi) e^{ikr}}{r},
\end{equation}

where $|f(\theta, \phi)|^2$ represents the differential scattering cross section $\sigma (\theta, \phi)$.

In the prolate spheroidal coordinate system, the plane wave can be expanded into a series of $S_{mn}(c, \eta)$ and $R_{mn}(c\xi)$:

\begin{equation}\label{physical expansion}
e^{i \bm{k} \cdot \bm{r}} = 2 \sum_{m,n} i^n \frac{(2 - \delta_{0m})}{N_{mn}} S_{mn}(c, \cos \theta_0) S_{mn}(c, \eta) R_{mn}^{(1)}(c\xi) \cos m (\phi - \phi_0).
\end{equation}

By invoking rotational symmetry, we can always orient the $x$ and $y$ axes such that $\phi_0 = 0$.

As $c\xi \to \infty$, $c\xi \to kr$, and $\eta \to \cos \theta$, the plane wave simplifies to:

\begin{equation}\label{physical_asym}
e^{i \bm{k} \cdot \bm{r}} = \sum_{m,n} i^n (2 - \delta_{0m}) \frac{(2n + 1) (n - m)!}{(n + m)!} P_n^m (\cos \theta_0) P_n^m (\cos \theta) \frac{\sin (kr - n\pi/2)}{kr} \cos m \phi.
\end{equation}

Substituting Eq. (\ref{physical_asym}) into Eq. (\ref{physical solution}), we obtain the asymptotic form of $\psi(\bm{r})$. By setting $\psi_{\text{mathematics}} = \psi_{\text{physics}}$ and equating the coefficients of $\exp(\pm ikr)$, we derive:

\begin{align}\label{f}
  f(\theta, \phi) &= f_{\theta_0} (\theta, \phi) \notag\\
  &= \sum_{mn} (2 - \delta_{0m}) \frac{(2n+1)(n-m)!}{(n+m)!} P_n^m (\cos \theta_0) P_n^m (\cos \theta) \cos m \phi \frac{e^{2i\gamma_{mn}} - 1}{2ik}.
\end{align}

The differential scattering cross section is given by:

\begin{equation}\label{d scattering}
\frac{1}{k^2} \sum_{mn} \left[ (2 - \delta_{0m}) \frac{(2n+1)(n-m)!}{(n+m)!} P_n^m (\cos \theta_0) \right]^2 \left[ P_n^m (\cos \theta) \cos m \phi \right]^2 \sin^2 \gamma_{mn}.
\end{equation}

The total scattering cross section can be obtained through integration:

\begin{align}\label{total section}
  \sigma_{\text{total} \, \theta_0} (\theta, \phi) &= \iint \sigma_{\theta_0} (\theta, \phi) \sin \theta \, d\theta \, d\phi \notag \\
   &= \frac{4\pi}{k^2} \sum_{mn} (2 - \delta_{0m}) \frac{(2n+1)(n-m)!}{(n+m)!} \left[ P_n^m (\cos \theta_0) \right]^2 \sin^2 \gamma_{mn}.
\end{align}

Let us consider the case of low-energy scattering, specifically when $c = \frac{1}{2} kd \to 0$ and $c\epsilon^{-1} \to 0$. In this scenario, for each given $m$ and $n$, the dominant term among all $d_r^{mn}$ is $d_{(n-m)}^{mn}$. Furthermore, we have:

\begin{subequations}\label{asymp of f}
\begin{align}
  \frac{d_{(n-m+2r)}^{mn}}{d_{(n-m)}^{mn}} & \approx c^{2r},\\
  \frac{j_{(n+2r)} (c\epsilon^{-1})}{j_n (c\epsilon^{-1})} & \approx \frac{(c\epsilon^{-1})^{2r}}{(2n+3) \cdots (2n+4r+1)}, \\
  \frac{n_{(n+2r)} (c\epsilon^{-1})}{n_n (c\epsilon^{-1})} & \approx \frac{(2n-3) \cdots (2n+4r-1)}{(c\epsilon^{-1})^{2r}}.
\end{align}
\end{subequations}

These imply that the relevant terms in the numerator include $d_{(n-m)}^{mn} j_n (c\epsilon^{-1})$ and all preceding terms in ascending order of $r$. Similarly, the significant terms in the denominator are $d_{(n-m)}^{mn} n_n (c\epsilon^{-1})$ and all subsequent terms in ascending order of $r$.

As $r \to \infty$, the ratio $\frac{d_{(2+r)}^{mn}}{d_r^{mn}}$ approaches zero at the rate of $-\frac{c^2}{4r^2}$, ensuring the series in the denominator converges. Thus, Eq. (\ref{gamma mn}) can be simplified to:

\begin{equation}\label{gamma mn simplified}
\tan (\gamma_{mn}) = \frac{j_n (c\epsilon^{-1})}{n_n (c\epsilon^{-1})} f_{mn} (\epsilon) \approx -\frac{(c\epsilon^{-1})^{2n+1}}{[1 \cdot 3 \cdot 5 \cdots (2n-1)]^2 (2n+1)} f_{mn} (\epsilon),
\end{equation}

where $f_{mn} (\epsilon)$ is of order 1:

\begin{equation}\label{fmn}
f_{mn}(\epsilon) = \frac{1+\sum\limits_{r\leq n-m-2}' i^{r} \frac{(2m+r)!(n-m)!}{(n+m)!r!}\frac{d_r^{mn}(c)}{d_{n-m}^{mn}} \frac{j_{m+r}(c\epsilon^{-1})}{j_n(c\epsilon^{-1})}}{1+\sum\limits_{r\geq n-m+2}' i^{r} \frac{(2m+r)!(n-m)!}{(n+m)!r!}\frac{d_r^{mn}(c)}{d_{n-m}^{mn}} \frac{n_{m+r}(c\epsilon^{-1})}{n_n(c\epsilon^{-1})}}.
\end{equation}

Thus, when computing the total scattering cross section, the dominant contribution arises solely from the term corresponding to $n = m = 0$:

\begin{equation}\label{total section final}
\tan (\gamma_{00}) = \frac{j_0 (c\epsilon^{-1})}{n_0 (c\epsilon^{-1})} f_{00} (\epsilon) \approx -c\epsilon^{-1} f_{00} (\epsilon).
\end{equation}

Substituting Eq. (\ref{total section final}) into Eq. (\ref{total section}) yields:

\begin{equation}
\sigma_{\theta_0\, \text{total}} (\theta, \phi) = \frac{4\pi}{k^2} (c\epsilon^{-1})^2 [f_{00} (\epsilon)]^2 = 4\pi a^2 [f_{00} (\epsilon)]^2,
\end{equation}

where $a = \frac{d}{2\epsilon}$ is the semi-major axis length.

The surface area of the ellipsoid is given by:

\begin{equation}\label{area}
S = 4\pi ab \int_0^\pi \sqrt{1 - \epsilon^2 \cos^2 \theta} \sin \theta \, d\theta = 4\pi a^2 g(\epsilon),
\end{equation}

where

\begin{equation}\label{g}
g(\epsilon)=\frac{1 - \epsilon^2 + c^{-1}(1 - \epsilon^2)^{1/2} \arcsin \epsilon}{2}.
\end{equation}

As we know, $f_{00}(\epsilon)$ is a value dependent on $c$ since $d_r^{mn}$ varies with $c$. We choose $c=0.1$, and $f_{00} (\epsilon)$ is expressed as:

\begin{equation}\label{f00}
f_{00} (\epsilon) = \frac{1}{1 + \frac{1}{3}\epsilon^2 + \frac{1}{5}\epsilon^4 + \frac{1}{7}\epsilon^6 + \cdots}.
\end{equation}

Therefore, up to the accuracy of $\epsilon^6$, we plot the following figure:

\begin{figure}[h!]
\centering
\includegraphics[width=\textwidth]{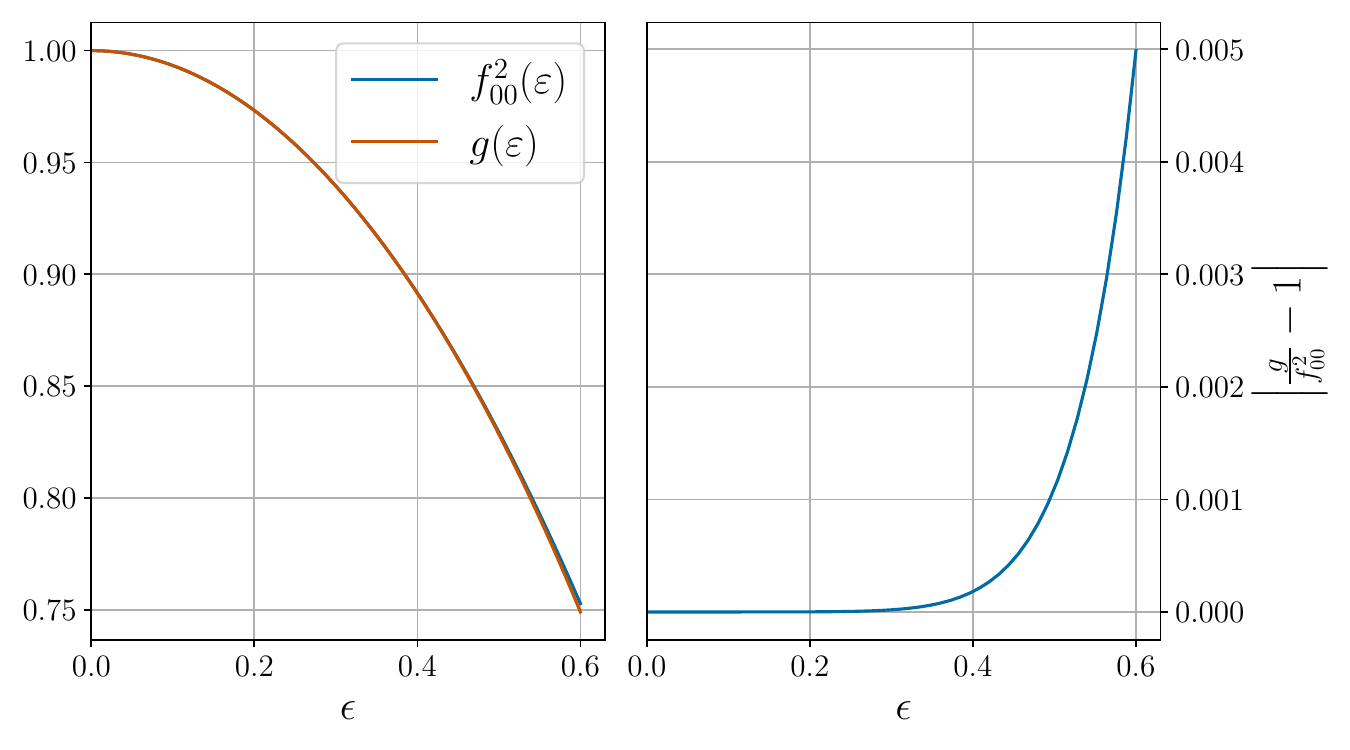}
\caption{Plot of $g(\epsilon), |f_{00} (\epsilon)|^2$ and relative error for prolate.}
\end{figure}

It can be observed that by performing the transformations in Eq. (\ref{trans}), the equations for the prolate spheroid can be transformed into those for the oblate spheroid. Consequently, by slightly modifying the results obtained above, we can derive the conclusions for the rotating oblate spheroid:

\begin{equation}\label{gamma1}
\tan(\gamma_{mn}) = \frac{R_{mn}^{(1)}(c\epsilon^{-1})}{R_{mn}^{(2)}(c\epsilon^{-1})} = \frac{\sum_{nm}' i^{(r+m-n)} d_r^{mn} (ic) \frac{(2m+r)!}{r!} j_{m+r}(c\epsilon^{-1})}{\sum_{nm}' i^{(r+m-n)} d_r^{mn} (ic) \frac{(2m+r)!}{r!} n_{m+r}(c\epsilon^{-1})},
\end{equation}

\begin{align}
\sigma_{\theta_0}(\theta, \phi) &= |f_{\theta_0} (\theta, \phi)|^2 \notag\\
&= \frac{1}{k^2} \sum_{mn} \left[(2-\delta_{0m})(2n+1)\frac{(n-m)!}{(n+m)!} P_n^m (\cos \theta_0) \right]^2 \left[P_n^m (\cos \theta) \cos m \phi \right]^2 \sin^2 \gamma_{mn} , \\
\sigma_{\text{total}\,\theta_0} (\theta, \phi) &= \int \sigma_{\theta_0} (\theta, \phi) \sin \theta \, d\theta \, d\phi \notag\\
&= \frac{4\pi}{k^2} \sum_{mn} (2-\delta_{0m})(2n+1)\frac{(n-m)!}{(n+m)!}\left[P_n^m (\cos \theta_0) \right]^2 \sin^2 \gamma_{mn}.
\end{align}\label{sigma}

In the case of low-energy scattering:

\begin{equation}\label{gamma1_asymp}
\tan (\gamma_{mn}) = \frac{j_n (c\epsilon^{-1})}{n_n (c\epsilon^{-1})} f_{mn} (\epsilon) \approx -\frac{(c\epsilon^{-1})^{2n+1}}{[1 \cdot 3 \cdot 5 \cdots (2n-1)]^2 (2n+1)} f_{mn} (i\epsilon),
\end{equation}

where the magnitude of \(f_{mn} (i\epsilon)\) is of order 1.

Thus, when calculating the total scattering cross-section, the primary contribution arises from the term corresponding to \(n = m = 0\):

\begin{equation}
\tan(\gamma_{00}) = \frac{j_0 (c\epsilon^{-1})}{n_0 (c\epsilon^{-1})} f_{00}(i\epsilon) \approx -c\epsilon^{-1} f_{00}(i\epsilon),
\end{equation}

yielding

\begin{equation}
\sigma_{\text{total}\,0} (\theta, \phi) = 4\pi a^2 [f_{00}(i\epsilon)]^2,
\end{equation}

where

\begin{equation}
f_{00}(i\epsilon) = \frac{1}{1 - \frac{1}{3}\epsilon^2 + \frac{1}{5}\epsilon^4 - \frac{1}{7}\epsilon^6 + \cdots},
\end{equation}

and \(a = \frac{d}{2\epsilon}\) is the length of the semi-minor axis.

The surface area of the ellipsoid is:

\begin{equation}
S = 4\pi ab \int_0^\pi \sqrt{1 + \epsilon^2 \cos^2 \theta} \sin \theta \, d\theta = 4\pi a^2 g(i\epsilon).
\end{equation}

We plot $g(i\epsilon), |f_{00}(i\epsilon)|^2$, and their relative error in the interval $\epsilon \in [0, 0.6]$:

\begin{figure}[h!]
\centering
\includegraphics[width=\textwidth]{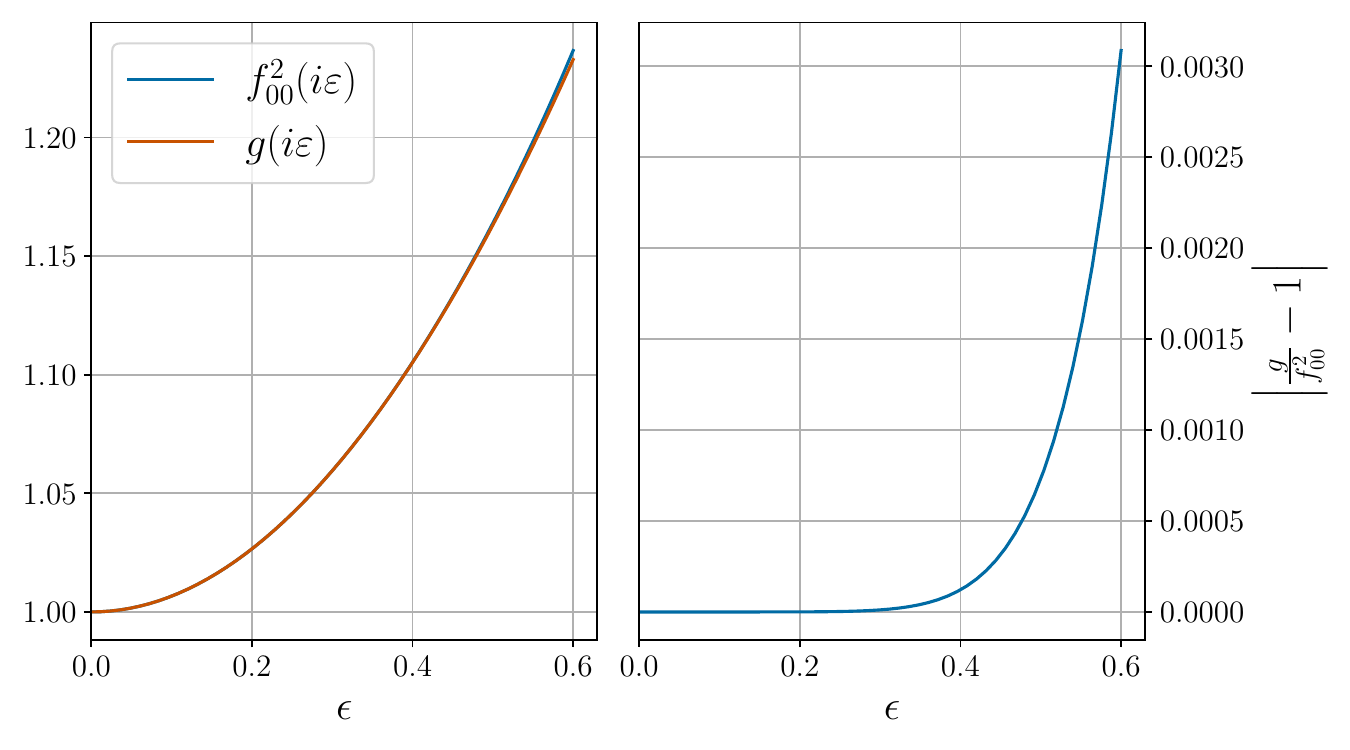}
\caption{Plot of $g(i\epsilon), |f_{00} (i\epsilon)|^2$ and their relative error for oblate.}
\end{figure}

\section{Conclusion}
In this study, we have demonstrated that under the long-wavelength approximation, the surface area of an ellipsoid closely approximates the total scattering cross-section, regardless of the direction of the incident wave. This holds true not only for small eccentricities but also for moderate values of eccentricity. This finding aligns with the established conclusions for spherical rigid bodies, suggesting a broader applicability of this principle.

We propose that the total scattering cross-section may be universally equivalent to the surface area of any scattering rigid body that the wave may reach, regardless of its geometric shape. While our results provide a compelling basis for this assertion, we acknowledge the need for a rigorous mathematical proof to substantiate this hypothesis. Future work will focus on developing such a proof, further elucidating the relationship between geometric properties and scattering behavior in various configurations.

\section*{Acknowledgments}
The author would like to thank for Chao Zhang for helpful discussion and kind help.
Y-N.Li is supported by National Natural Science Foundation of China under Grant No. 12005295.

\bibliographystyle{apsrev4-1}
\bibliography{Scatering}
\end{document}